\documentclass[11pt]{article}
\usepackage{amssymb}
\usepackage{latexsym}
\usepackage[mathscr]{eucal}
\usepackage{citesort}
\usepackage{url}
\usepackage{ntheorem}
\newcommand{\mcNkT}{\mcN_{\vec k,T}}
\newcommand{\doc}{\lessdot\Mext\gtrdot}
\newcommand{\Gp}{\Gamma^+}
\newcommand{\Gm}{\Gamma^-}

\newcommand{\gm}{\gamma^-}

\newcommand{\beaa}{\begin{eqnarray*}}
\newcommand{\eeaa}{\end{eqnarray*}}
\newcommand{\bean}{\begin{eqnarray}\nonumber}%
\newcommand{\GtRpmn}{\Gamma_{p}^\pm}%
\newcommand{\GtRpn}{\Gamma_{p}^+}%
\newcommand{\GzRp}{\Gp}%
\newcommand{\GtRmn}{\Gamma_{p}^-}%
\newcommand{\GzRm}{\Gm}%
\newcommand{\gzRm}{\gm}%

\newcommand{\Mext}{\mcM_\ext}

\newcommand{\hmcN}{\,\,\,\,\tilde{\!\!\!\!{\mycal N}}}%
\newcommand{\mcN}{{\mycal N}}%
\newcommand{\mcP}{{\mycal P}}%
\newcommand{\zGamma}{\mathring \Gamma}%
\newcommand{\zG}{\mathring \Gamma}%
\newcommand{\zR}{\mathring R}%
\newcommand{\zC}{\mathring C}%
\newcommand{\ext}{{\mbox \scriptsize \rm ext}}

\newcounter{mnotecount}[section]

\newcommand{\id}{{\mbox{\rm Id}}}

\newcommand{\mcM}{{\mycal M}}

\newcommand{\hyp}{{\mycal S}}

\def \Reel{\mathbb{R}}
\def \R {\Reel}

\def \Nat{\mathbb{N}}

\def \N {\Nat}

\newcommand{\be}{\begin{equation}}
\newcommand{\ee}{\end{equation}}
\newcommand{\bel}[1]{\begin{equation}\label{#1}}
\newcommand{\beal}[1]{\begin{eqnarray}\label{#1}}
\newcommand{\beadl}[1]{\begin{deqarr}\label{#1}}
\newcommand{\eeadl}[1]{\arrlabel{#1}\end{deqarr}}
\newcommand{\eeal}[1]{\label{#1}\end{eqnarray}}
\newcommand{\eead}[1]{\end{deqarr}}
\newcommand{\eea}{\end{eqnarray}}

\newcommand{\Hess}{\mathrm{Hess}\,}

\newcommand{\eq}[1]{(\ref{#1})}
\newcommand{\Eq}[1]{Equation~(\ref{#1})}

\DeclareFontFamily{OT1}{rsfs}{}
\DeclareFontShape{OT1}{rsfs}{m}{n}{ <-7> rsfs5 <7-10> rsfs7 <10->
rsfs10}{} \DeclareMathAlphabet{\mycal}{OT1}{rsfs}{m}{n}

\newcommand{\mcD}{{\mycal D}}

\newcommand{\mcO}{{\mycal O}}
\newcommand{\mcT}{{\mycal T}}
\newcommand{\mcU}{{\mycal U}}

\newcommand{\qed}{\hfill $\Box$\bigskip}

\newcommand{\proof}{\noindent {\sc Proof:\ }}

\newtheorem{defi}{\sc Definition\rm}[section]

\newtheorem{Theorem}[defi]{\sc Theorem\rm}

\newtheorem{Proposition}[defi]{\sc Proposition\rm}

\newtheorem{Lemma}[defi]{\sc Lemma\rm}

\newtheorem{Corollary}[defi]{\sc Corollary\rm}

\theoremstyle{remark} \theorembodyfont{\upshape}
\newtheorem{Remark}[defi]{\sc Remark\rm}

{\catcode `\@=11 \global\let\AddToReset=\@addtoreset}
\AddToReset{equation}{section}
\renewcommand{\theequation}{\thesection.\arabic{equation}}

\begin{document}
\title{A poor man's positive energy theorem: II. Null geodesics}
\author{
Piotr T. Chru\'sciel\thanks{Partially supported by a Polish
Research Committee grant 2 P03B 073 24; email
\protect\url{Piotr.Chrusciel@lmpt.univ-tours.fr}, URL
\protect\url{ www.phys.univ-tours.fr/}$\sim$\protect\url{piotr}}\\
D\'epartement de
Math\'ematiques\\
Facult\'e des Sciences\\ Parc de Grandmont\\ F37200 Tours, France
}
\date{}

\maketitle

\begin{abstract}
We show that positivity of energy for stationary, or
\emph{strongly uniformly Schwarzschildian}, asymptotically flat,
non-singular domains of outer communications can be proved using
Galloway's null rigidity theorem.
\end{abstract}


\section{Introduction}

 In a recent
note~\cite{poorman} we have shown that positivity of energy for
uniformly Schwarzschildian, asymptotically flat, non-singular
domains of outer communications can be proved using timelike lines
together with the Lorentzian splitting theorem. The object of this
paper is to show that 
a variation of the argument of~\cite{poorman}, using null lines,
can successfully be completed in a similar, ``strongly uniformly
Schwarzschildian", setting. The extension of the result to more
general metrics, as suggested in~\cite{PenroseSorkinWoolgar},
still awaits  a more complete justification.

For $m\in\R$, let $g_m$ denote the $n+1$ dimensional, $n\ge 3$,
Schwarzschild metric with mass parameter $m$; in isotropic
coordinates~\cite{Patel:1999ej}, \bel{stschw} g_m= \left(1 +
\frac{m}{2|x|^{n-2}}\right)^{\frac4{n-2}} \left(\sum_{1=1}^n
dx_i^2\right) - \left(\frac{1-m/2|x|^{n-2}}{1+m/2|x|^{n-2}}\right)^2
dt^2\;. \ee We shall say that a metric $g$ on $\R\times
\left(\R^n\setminus B(0,R)\right)$, $R^{n-2}>m/2$, is \emph{strongly
uniformly Schwarzchildian} if, in the coordinates of \eq{stschw},
\beal{guS1n} &g-g_m = O(|m|r^{-(n-1)})\;,\quad
\partial_\mu\left(g-g_m \right)= O(|m|r^{-n})\;, & \\ & \partial_\mu\partial_\nu\left(g-g_m \right)
= O(|m|r^{-n-1})\;.\eeal{guS2n} (Here $O$ is meant at fixed $g$ and
$m$, uniformly in $t$ and in angular variables, with $r$ going to
infinity.) What is meant in the case $m=0$ is that $g= g_0$, i.e., g
is flat\footnote{The asymptotic conditions for the case $m=0$ of our
theorem are way too strong for a rigidity statement of real
interest, even for stationary metrics, so our results only exclude
$m<0$ in practice. Nevertheless, in this context one should keep in
mind an argument of Lohkamp~\cite{Lohkamp} for initial data sets
with trace of $K=0$,  which shows that the proof of positivity of
mass is obtained in its full generality once it has been established
for data which are flat outside of a compact set.}, for $r>R$. We
note that this condition is more restrictive than the ``uniformly
Schwarzschildian" one in~\cite{poorman}.

We will use the symbol $r$ to denote some fixed smooth function on
$\mcM$ which coincides with $|x|$ in $\Mext$.

The \emph{domain of outer communications} $\doc$ associated to
$\Mext$ is defined
  as the
intersection of the causal past $J^-(\mcM_\ext)$ of the asymptotic
region
\bel{mext} \mcM_\ext=\R\times \left(\R^n\setminus
B(0,R)\right)\ee with its causal future $J^+(\mcM_\ext)$.

We need a version of weak asymptotic simplicity~\cite{HE} for
uniformly Schwarzschildian spacetimes.  Following~\cite{poorman},
we shall say that such a spacetime $(\mcM,g)$ is \emph{weakly
asymptotically regular}  if every null line starting in the domain
of outer communications $\doc$ either crosses an event horizon (if
any), or reaches arbitrarily large values of $r$ in the
asymptotically flat regions.  Recall that a null line in
$(\mcM,g)$ is an inextendible null geodesic that is globally
achronal. 

The first main result of this paper is a simple proof of the
following:

\begin{Theorem}
\label{Tpoormannull} Let $(\mcM^{n+1}=\mcM,g)$ be a
$(n+1)$-dimensional  weakly asymptotically regular space-time
containing a strongly uniformly  Schwarzschildian region $\Mext$
 with $$m\le 0\;.$$  If the
domain of outer communications associated to $\mcM_\ext$ has a
Cauchy surface $\hyp$, the closure of which is the union of one
asymptotic end and of a compact interior region (with a
differentiable boundary lying at the intersection of the future and
past event horizons, if any), then $\mcM$ contains a null line.
\end{Theorem}

 Theorem~\ref{Tpoormannull}
implies non-existence of appropriately regular, uniformly
Schwarzschildian space-times with negative mass in various
situations of interest. For example, consider a uniformly
Schwarzschildian four-dimensional, asymptotically simple, vacuum
space-time $(\mcM,g)$ with $m\le 0$. It then follows from
Theorem~\ref{Tpoormannull} and from Galloway's null rigidity
theorem~\cite{Galloway:splitting} that $(\mcM,g)$ is the Minkowski
space-time. This result is somewhat stronger than the one
in~\cite{poorman}, because no hypotheses on timelike geodesics are
made (we will show below that, similarly to~\cite{poorman}, in our
context the hypothesis of asymptotic simplicity needed
in~\cite{Galloway:splitting} can be replaced by the weak version
thereof). On the other hand, there is a restriction on space-time
dimension, not made in~\cite{poorman}.

The original idea in~\cite{Penrose:nmpt} was  that space-times
containing null lines cannot satisfy the so-called ``genericity
condition". However, 
it is not clear how generic is ``generic" when, e.g., vacuum
equations are imposed, and therefore, it is not clear how much
information is carried by theorems basing upon ``genericity". For
this reason it appears useful to develop an argument which does
not invoke this condition. As a first step towards this, assuming
the \emph{null energy condition}, \bel{ncc} R_{\mu\nu}X^\mu X^\nu
\ge 0 \ \mbox{for all null vectors $X^\mu$,}\ee we prove the
following:

\begin{Theorem}
\label{Tfoliate} Under the hypotheses of Theorem~\ref{Tpoormannull},
suppose moreover that the null energy condition~\eq{ncc} holds. Then
$$\mcM=\doc\;,$$ and every maximally extended null geodesic is a
line. Further,
 for each $\vec k \in S^{n-2}$ there exists a one-parameter
family $\mcN_{\vec k}\equiv\{\mcN_{\vec k,T}\}_{T\in\R}$, with
$\mcN_{\vec k'}\ne \mcN_{\vec k}$ if $\vec k\ne \vec k'$, of
smooth, closed, null, totally geodesic, achronal hypersurfaces
covering $\mcM$.
\end{Theorem}


With a little more effort one can show that each family
$\mcN_{\vec k}$ forms a foliation; we give no details as we will
not pursue this line of approach. Now, each foliation of $\mcM$ by
null hypersurfaces as described in Theorem~\ref{Tfoliate} defines
a shear-free, rotation-free, and expansion-free congruence of null
geodesics covering $\mcM$. Space-times admitting one such
congruence are expected to be very special\footnote{See,
e.g.,~\cite[Section~7.6 and Chapter~31]{Exactsolutions2}. The
results there are based upon a supplementary assumption on the
vanishing of some components of the Ricci tensor. Note, however,
that it follows from the Raychaudhuri equation that this condition
will be satisfied in our case if the dominant energy condition is
assumed.}, while we actually have a family of congruences
parameterized by $S^{n-1}$. We therefore expect that the only
space-time satisfying the hypotheses of Theorem~\ref{Tfoliate} is
the Minkowski one, but we are not aware any results to this effect
without further hypotheses (compare~\cite{Mason:AlgSpec}).
Assuming $n$ equal to four, as well as existence of a smooth
conformal completion,  we will be able to reach this conclusion by
an argument that does not proceed via the algebraically special
structure of the Riemann tensor; this is the main result of this
paper:

\begin{Theorem} \label{Tpoormannulnew} Let
$(\mcM^{3+1}=\mcM,g)$ be a four dimensional space-time satisfying
the null energy condition \eq{ncc}, and suppose that $\mcM$
contains a strongly uniformly Schwarzschildian region
$\mcM_\ext=\R\times \left(\R^3\setminus B(0,R)\right)$ which
admits a smooth conformal completion at null infinity. Assume that
$(\mcM,g)$ is weakly asymptotically regular, and that the domain
of outer communications of $\mcM$ has a Cauchy surface $\hyp$, the
closure of which is the union of one asymptotic end and of a
compact interior region (with a differentiable  boundary lying at
the intersection of the future and past event horizons, if any).
Then
$$m>0\;,$$ unless $(\mcM,g)$ is the Minkowski space-time.
\end{Theorem}

\medskip

It is of interest to compare the results and proofs here to those
in~\cite{poorman}. Recall that we are trying to implement  the
original idea of Penrose~\cite{Penrose:nmpt}, that negative mass
should imply existence of a \emph{null line} in  space-time, which
then should be incompatible with energy conditions.
In~\cite{poorman} we have shown, under the hypotheses there, that
negative mass implies existence of a \emph{timelike line}, and
that this is compatible with the timelike convergence condition,
\bel{tcc} R_{\mu\nu}X^\mu X^\nu \ge 0 \ \mbox{for all timelike
vectors $X^\mu$,}\ee only if the space-time is the Minkowski one.
Both in~\cite{poorman} and here we make the unsatisfactory
hypothesis that the metric is uniformly Schwarzschildian. This is,
however, a quite reasonable hypothesis for \emph{stationary}
space-times with $m\ne 0$ (see, e.g.,~\cite{Simon:1984kz}). In the
stationary case the only global hypothesis needed for both
constructions is that of existence of an asymptotically flat
Cauchy surface for the domain of outer communications, with
conditionally compact interior. For stationary metrics the proof
of existence of lines is completely elementary for timelike lines
in~\cite{poorman}, and marginally more complicated for null lines
here.

It is not clear at all whether there exist \emph{non-stationary}
solutions of physically reasonable field equations that are
uniformly Schwarzschildian, and therefore assuming that last
condition without assuming stationarity is presumably an academic
exercise. We have nevertheless carried this out, in order to test
the limits of the techniques employed, and in the hope that the
arguments  will eventually generalise to a proof under less
stringent asymptotic conditions. As already pointed out, in the
non-stationary case the construction of a null line is only
slightly more involved than that of a timelike one. On the other
hand, the conclusion that the space-time must be flat requires
considerably more work. However, the proof of existence of a
timelike line needs a supplementary global hypothesis concerning
timelike geodesics, which is avoided in the current setting. This
last fact provides yet another motivation for the work here.

The next step in the positivity argument is to show that
space-times containing lines and satisfying energy conditions are
very special. In considerations involving timelike lines the
natural energy condition is the unphysical inequality \eq{tcc}.
{}From this point of view null lines are much more satisfactory,
as their global causality properties are tied to the physically
motivated \emph{null energy condition} \eq{ncc}. This is the final
motivation for our work.

This paper is organised as follows:  In Section~\ref{Sproof} we
give the proofs of our results. Appendix~\ref{SA} contains some
technical results on geodesics in uniformly Schwarzschildian
metrics, as needed in the proofs of Theorems~\ref{Tfoliate} and
\ref{Tpoormannulnew}.

\section{Proofs}
\label{Sproof} We start with a proof of
Theorem~\ref{Tpoormannull}. We will be sketchy in places, assuming
that the reader is familiar with the argument in~\cite{poorman}.

Let $x^\mu$ be the coordinates of \eq{stschw} ($x^0=t$),
and let the indices $a,b,..$ run from $1$ to $n-1$, where $n$ is
the space-dimension. Using \eq{Gam1}-\eq{Gam2} in the appendix
below, one finds that the function
$$\rho:= \sqrt{\sum_{a=1}^{n-1}(x^a)^2}$$
has convexity properties somewhat similar to those of $r$, as
exploited in~\cite{poorman}: \bean\Hess \rho &=&
-\frac{m\rho}{r^n}\left\{(n-2)dt^2 + \sum_{i=1}^n (dx^i)^2\right\}
+\frac{2m}{r^{n-1}}\  d\rho dr \\ && + \frac 1 \rho
\left\{\sum_{a=1}^{n-1} (dx^a)^2 -d\rho^2\right\} + \sum
_{\mu,\nu=0}^n\sum_{a=1}^{n-1} \frac {x^a}\rho O(r^{-n})dx^\mu
dx^\nu\;.\eeal{hessrho}

On level sets of $\rho$ all terms involving $d\rho$ drop out, then
both expressions in braces are manifestly positive for negative
$m$. In order to show that they dominate all the remaining terms
when $\rho\ge \mathring R$ for some $\mathring R$, one needs to
check that \bel{erro} \frac{|m|\rho}{r^n}\ge c(n)\frac {x^a}\rho
O\left(\frac 1 {r^n}\right) \ \Longleftrightarrow \ |m|\rho^2 \ge
c(n) x^a O(1)\;,\ee where $c(n)$ is a (large) dimension-dependent
constant. \Eq{erro} will clearly hold for $\rho$ large enough if
$m\ne 0$.
 It follows
 that $\Hess \rho$,
when restricted to $\{\rho=R\}$, with $R\ge\mathring R$, is
positive definite when $m<0$. This shows, as in~\cite{poorman},
for negative or vanishing $m$, that any geodesic segment $\Gamma$
with initial point $p$ and final point $q$ such that $\rho(p),
\rho(q)< \mathring R$ will satisfy \bel{geocond1}\rho \circ \Gamma
< \mathring R\;.\ee

We start with a lemma\footnote{I am grateful to G.Galloway for
providing this simple proof.}:
\begin{Lemma}\label{Lcocon} For every $p\in J^-(\Mext)$ there
exists
an achronal, future inextendible, null geodesic ray
$$\GtRpn:[0,\infty)\to J^-(\Mext)$$ such that
$$\GtRpn(0)=p\;, \qquad \rho\circ(\GtRpn)\le
M\;,$$ for some constant $M=M(p)$.
\end{Lemma}

\proof Since $p\in J^-(\Mext)$ for any $N\in\N$ there exists a
causal curve
from $p$ to a point 
\bel{pform} p^{+}_{ N}=(t^+_N,0,\cdots,0, N)\in\Mext\;,\ee for some
$t^+_N$. Minimising the time of arrival $t^+_N$ over all such
causal curves one obtains, by global hyperbolicity, a null
achronal geodesic segment $\Gamma_N^+$ from $p$ to perhaps a
different point $p^{+}_{ N}$, still  of the form \eq{pform}. By
the convexity properties of the function $\rho$, discussed above,
we have the bound $\rho \circ \Gamma^+_N \le M$. Global
hyperbolicity implies that there exists a subsequence of the
sequence $\Gamma_N$ accumulating at the desired inextendible null
geodesic ray $\GtRpn$. Achronality of $\GtRpn$ follows from the
fact that an accumulation curve of achronal curves is achronal.
\qed

It immediately follows that:

\begin{Corollary}
\label{Ccocon}For every $p\in
\doc$ there exist achronal, future inextendible, null geodesic rays
$$\GtRpn:[0,\infty)\to\doc\ \mbox{ and } \
\GtRmn:(-\infty,0]\to\doc$$ such that
$$\GtRpmn(0)=p\;, \qquad \rho\circ(\GtRpmn)\le
M\;,$$ for some constant $M=M(p)$.
\end{Corollary}

Choose $\zR$ large enough so that the hypersurfaces $\{x^n=\pm
\zR\}$ are closed boundaryless in $\Mext$. Increasing $\zR$ we can
also assume that $\partial_t$ is timelike for $r\ge \zR$, that the
slopes of the light cones in the region $r\ge \zR$ are between
$1/2$ and 2, and that $\Hess \rho$ is positive definite on the
level sets $\{\rho=R\}$ for all $R\ge \zR$.

Choose $\tau\in\R$, let $p=(\tau,0,\ldots,0,-\zR)$, set
$$\Gp:=\Gamma_{p}^+\;, \quad \Gm:=\Gamma_{p}^-\;,$$ and define
$$\mcN^+:= \dot J^-(\Gp )\;,\quad  \mcN^-:= \dot J^+(\Gm)\;.$$
Since the $\Gamma^\pm$'s are achronal we have
$\Gamma^\pm\subset\mcN^\pm$, thus the $\mcN^\pm$'s are nonempty,
closed, achronal, locally Lipschitz submanifolds of $\mcM$. Denote
by
$$\mcP^\pm := \mcN^\pm \cap \{x^n=-\zR\}\;.$$
The manifolds $\{x^0=\pm \zR\}$ with the induced metric are globally
hyperbolic submanifolds of $(\mcM,g)$, with the $\mcP^\pm$ non-empty
(as $\Gamma^{\pm}\cap \{r=\pm \zR\}=p\in \mcP^\pm$), which shows
that the $\mcP^\pm$'s are closed, achronal, locally Lipschitz
submanifolds of $\{x^n=-\zR\}$, and can thus be written as graphs
$$\mcP^\pm = \{t=s^\pm (x^a)\}\;,$$ for some locally Lipschitz
functions $s^\pm$. 

 In the coordinates of \eq{stschw} the null geodesics
$\Gamma^\pm$ can be written as
$$\Gamma^\pm(s)=(t^\pm(s), \gamma^\pm(s))\;.$$
For $k\in\N$ let $p_k =\GzRp(k)$, set
$$q_k= \dot J^-(p_k)\cap \left(\R\times\{\gzRm(-k)\}\right)\;.$$
By global hyperbolicity there exists a null achronal geodesic
$\Gamma_k$ from $q_k$ to $p_k$. By Corollary~\ref{Ccocon} there
exists a constant $M$ such that $\rho(q_k),\rho(p_k)\le M$, and by
the argument presented in~\cite{poorman} we then have $\rho\circ
\Gamma_k \le M$ for all $k$ large enough. Let
$$r_k= \Gamma_k\cap \{x^n=-\zR\}\;.$$ We claim that $r_k$ satisfies \bel{rksat} r_k \in \underbrace{\{(x^0,x^a):\
s^-(x^a)\le x^0 \le s^+(x^a)\;,\ \rho(x^a)\le M\}}_{=:K}\subset
\{x^n=-\zR\}\;.\ee The upper bound on $x^0$ follows immediately from
$r_k\in J^-(\GzRp)$. To obtain the lower one we show that $r_k\in
J^+(\GzRm)$: Indeed, since achronal null geodesics maximise the time
of arrival (see, e.g.,~\cite[Proposition~A.2]{Chrusciel:2002mi}) we
have
$$t(q_k)=\sup_{p\in J^-(p_k)\cap \{\vec x = \gzRm(-k)\}}t(p)\ge t^-(-k)=t(\GzRm(-k))\;.$$
This shows that a future directed causal curve from $\Gamma^-(-k)$
to $r_k$ is obtained by first following the coordinate line
$$[t^-(-k),t(r_k)]\ni t\to (t,\gzRm(-k))\;,$$ and then following $\Gamma_k$ until it meets $\{x^n=-\zR\}$.

We have thus shown that all the achronal geodesic segments
$\Gamma_k$ pass through the compact set $K$ defined in \eq{rksat},
and the existence of a null line $\Gamma$ at which the
$\Gamma_k$'s accumulate follows from standard results in
Lorentzian geometry. \qed

\medskip

 {\noindent \sc Proof of
Theorem~\ref{Tfoliate}:} Let $\Gamma$ be the line constructed in
the proof of Theorem~\ref{Tpoormannull}. Writing
 $\Gamma\cap \Mext$ in coordinates on $\Mext$ as
$x^\mu(s)$, by Corollary~\ref{Cuniq} there exists $k$, $T$ and
$\beta$ such that \eq{alpf}-\eq{desform} hold. Since $\rho$ is
bounded along $\Gamma$ we must have $k=(1,0,\ldots,0,1)$.

The construction of $\Gamma$ applies for every choice of asymptotic
direction $k=(1, \vec k)$, we shall denote by $\zGamma_{\vec k, T}$
the resulting line. By Corollary~\ref{Cuniq} we have \bel{coinc}
\zGamma_{\vec k, T}=\Gamma_{(1,\vec k), T,\beta(\vec k,T)}\ee for
some $\beta(\vec k,T)\in \R^{n-1}$. Note that $\Gamma$ depends upon
a parameter $\tau\in\R$, so this equation defines a function
$T(\tau)$. The construction of $\Gamma$, together with
Corollary~\ref{Cuniq}, similarly shows that
\bel{symmetry}\zGamma_{\vec k, T}=\Gamma_{(-1,-\vec k),
T',\beta'(\vec k,T)}\;,\ee for some $T'\in \R$ and $\beta'\in
\R^{n-1}$. Here $\R^{n-1}$ is understood as the plane orthogonal
to $\vec k$ with respect to the Euclidean metric on $\R^n$.

Let $\hmcN_{\vec k,T}$ denote the collection of points  in $\dot
J^-({\zGamma_{\vec k, T}})\cap \dot J^+({\zGamma_{\vec k, T}})$
 which lie on a generator of $\dot
J^-({\zGamma_{\vec k, T}})$  which is past complete,  future
complete, and entirely contained in $\doc$. Then $\hmcN_{\vec
k,T}$ is non-empty, as it contains $\zGamma_{\vec k, T}$. We
define
$$\mcN_{\vec k,T}=\{\mbox{the connected component of $\hmcN_{\vec k,T}$
containing  $\zGamma_{\vec k, T}$}\}.$$
 Assuming the null energy condition,  we claim that $\mcN_{\vec k,T}$
is a smooth null hypersurface. This
 results from the discussion in~\cite{Galloway:splitting}, and
 can be seen as follows: Let $p\in \zGamma_{\vec k,
 T}$, as $\Mext$ is open there exists a  neighborhood $\mcO$ of $p$ entirely contained
 in $\Mext$. Consider the generators of $\dot
J^-({\zGamma_{\vec k, T}})$ passing through $\mcO$. As $p$ is an
interior point of such a generator, all the null tangents in $T
\mcO$ can be made as close to the null tangent at $p$ as desired
if $\mcO$ is made small enough (see, e.g.,~\cite{ChGalloway}). It
follows from continuous dependence of geodesics upon initial data
that all those generators have asymptotic behavior as described in
Proposition~\ref{Pexg2} when $\mcO$ is chosen small enough (with
perhaps an asymptotic direction vector $\vec k'\ne \vec k$), in
particular they will be future complete. By Lemma~4.2 in
\cite{Galloway:splitting} the divergence of $\dot
J^-({\zGamma_{\vec k, T}})$ (as measured towards the future,
defined in the sense of support hypersurfaces) is non-negative at
those points of $\dot J^-({\zGamma_{\vec k, T}})$ which lie on
generators meeting $\mcO$. Similarly, reducing $\mcO$ if
necessary, the divergence of generators of $\dot
J^+({\zGamma_{\vec k, T}})$ (as measured again towards the future)
is non-positive at those points of $\dot J^-({\zGamma_{\vec k,
T}})$ which lie on generators meeting $\mcO$. By Theorem~3.4
of~\cite{Galloway:splitting} $\dot J^-({\zGamma_{\vec k, T}})\cap
\mcO$ is a smooth null hypersurface which coincides with $\dot
J^+({\zGamma_{\vec k, T}})\cap \mcO$. It is then easily seen that
the collection of points of $\dot J^-({\zGamma_{\vec k, T}})$
which lie on generators meeting $\mcO$ is a smooth null
hypersurface contained in $\dot J^+({\zGamma_{\vec k, T}})$, with
all the generators there being both future and past complete,
contained in $\Mext$, extending arbitrary far in the asymptotic
region both to the future and to the past.

We have thus shown that $\mcN_{\vec k,T}$ has all the properties
listed in the statement of the theorem except perhaps for being
closed. In order to establish that last property, consider a
sequence of points $p_n\in \mcNkT$ converging to $p\in \mcM$, and
let $\Gamma_n$ denote the generator of $\dot J^-({\zGamma_{\vec k,
T}})$ passing through $p_n$. Set
$$q_n=\Gamma_n\cap \hyp\;,$$ where $\hyp$ is the Cauchy surface
for $\doc$, then a subsequence, still denoted by $q_n$, converges to
a point $q\in \overline\hyp$. Let $\Gamma$ denote an accumulation
curve of the $\Gamma_n$'s passing through $q$, then $\Gamma$ is an
achronal, maximally extended in $\overline{\doc}$, geodesic passing
through $q$. Suppose that $q\in
\partial \hyp$, then the portion of $\Gamma_n$ which lies to the
future of $\hyp$ accumulates to a generator of
$\partial\mcD^+(\hyp)$, while the portion of $\Gamma_n$ which lies
to the past of $\hyp$ accumulates to a generator of
$\partial\mcD^-(\hyp)$, resulting in an accumulation curve
$\Gamma$ which is not differentiable at $q$. This, however,
contradicts the fact that $\Gamma$ is a geodesic. Hence $q\in
\hyp$, and $\Gamma$ is both future and past complete. It follows
that $q\in\mcNkT$, with $p$ lying on $\Gamma$. 
This shows that $p\in\mcNkT$, and closedness of $\mcNkT$ follows.

Note that so far $T$ was an arbitrarily chosen fixed number. If
the domain of outer communications is stationary one can move
$\mcN_{\vec k,T}$ by isometries to obtain a one-parameter family
$\mcN_{\vec k,t}:=\phi_{t-T}(\mcN_{\vec k, T})$, $t\in \R$, of
such hypersurfaces. A similar construction can be done in the
stationary-rotating case.

 In general, we start by noting that:

\begin{Lemma}
\label{Lmon3} Let $s_0\in \R$, $\vec k \in S^{n-1}$, $a=1,2$. For
$T_a \in \R$  and $\beta_a \in \R^{n-1}$ set
$\Gamma_a=\Gamma_{(1,\vec k),T_a,\beta_a}\big|_{[s_0,\infty)}$.
Then
$$T_1< T_2 \quad  \Longrightarrow \quad
\Gamma_1\subset I^-({\Gamma_2};\Mext)\;.$$ 
\end{Lemma}

\proof We write $x^\mu_a$ for $\Gamma_a$ in the asymptotically
Minkowskian coordinates on $\Mext$, and use an affine parameter as
in \eq{desform}. By Proposition~\ref{Puniq} for any $\lambda>0$ we
have 
\bel{gvec} x_2^\mu(s+\lambda)-x_1^\mu(s)\to_{s\to
\infty}(T_2-T_1+\lambda,\beta_2-\beta_1,\lambda)\;.\ee Consider
the coordinate-line segment
$$[0,1]\in t \to \gamma^\mu_s(t)= tx^\mu_2(s+\lambda) +
(1-t)x^\mu_1(s)\;.
$$
Calculating with respect to the Minkowski metric $\eta$, the
quantity $\eta_{\mu\nu}\dot\gamma^\mu_s \dot \gamma^\nu_s$ (a dot
denotes a $t$-derivative) approaches, as $s$ tends to infinity,
the Minkowskian length of the vector appearing at the
right-hand-side of \eq{gvec}, which is
$$-(T_2-T_1+\lambda)^2+|\beta_2-\beta_1|_\delta^2 + \lambda^2 =
-2(T_2-T_1)\lambda +|\beta_2-\beta_1|_\delta^2\;.$$ When $T_2>T_1$
this is negative for all $\lambda$ sufficiently large. It follows
that $\gamma_s$ is a timelike curve with respect to the Minkowski
metric for all $\lambda $ and $s$ large.  Since the metric $g$
uniformly tends to the Minkowski one along $\gamma_s$ as $s$ tends
to infinity, $\gamma_s$ will be a timelike curve with respect to
$g$ for all $\lambda $ and $s$ large enough. \qed

In what follows $\zR$ is chosen at least as large as in the proof
of Theorem~\ref{Tpoormannull}, and in \eq{Sminf}:

\begin{Corollary}
\label{Cmon} Let $T',\beta'$ be such that the geodesic $\GtRpn$,
$p=(\tau,0,\ldots,0,-\zR)$, of Lemma~\ref{Lcocon} is a subset of
$\Gamma_{(1,\vec k),T', \beta'}$ (when both are understood as point
sets in $\mcM$), and let $T=T(\tau)$ be as defined by \eq{coinc}.
Then $$T'=T\;.$$
\end{Corollary}
\proof Clearly $T\le T'$ by construction. Suppose $T<T'$, then
$\zGamma_{\vec k, T}$ would lie to the timelike past of $\GtRpn$
by Lemma~\ref{Lmon3}. But, by construction,  $\zGamma_{\vec k,
T}\subset \dot J^-(\GtRpn)$, whence the result. \qed

\begin{Corollary}
\label{Lmon} The map which to the parameter $\tau$  assigns $T$, as
defined by \eq{coinc}, is a diffeomorphism from $\R$ to $\R$.
\end{Corollary}

\proof The map $\tau\to T'(\tau)$ is the inverse of the map $\Phi$
defined after \Eq{Phimap}, Appendix~\ref{SA}, which is shown to be
a diffeomorphism of $\R$ there, and the result follows from
Corollary~\ref{Cmon}. \qed

We continue with

\begin{Proposition} \label{Lmon2} 
$\mcN_{\vec k,T}=\cup_{\beta\in \R^n}\Gamma_{(1,\vec k),T,\beta}$.
\end{Proposition}

\proof
 Let $p\in \mcN_{\vec k,T}$, then there exists a causal line
$\Gamma_p$ passing through $p$ contained in $\doc$. By weak
asymptotic regularity $\Gamma_p$ satisfies \eq{extend}, and thus
$\Gamma_p=\Gamma_{(1,\vec
 k'),T',\beta'}$ for some $\Gamma_{(1,\vec
 k'),T',\beta'}$ by Corollary~\ref{Cuniq}.

 We want, first, to show that
\bel{keq}\vec k'=\vec k\;.\ee Since $\mcNkT$ is a smooth
hypersurface its field of null tangents is also smooth. This,
together with connectedness of $\mcNkT$, allows one to reduce the
proof of \eq{keq} to a situation where $\vec k'$ is close to $\vec
k$. We shall therefore assume that $\vec k=(0,\ldots,0,1)$ and
that $ k'^n>\sqrt {15} / 4$. It follows from \eq{desform} that
choosing $\zR$ and $s_0$ large enough one can also assume that
$\Gamma_{(1,\vec
 k'),T',\beta'}(s)\in \{x^n\ge \zR\}$ for $s\ge s_0$.

Let $s$ be an affine parameter as in \eq{alpf}--\eq{decest2}, and
for $i\in\N$ let $p_i=p_i(s)\in \Mext$ denote a point with
coordinates $(x^0(s)-1/i,\vec x(s))$. Since $p_i\in
I^-(\Gamma_{(1,\vec
 k'),T',\beta'})$ we have $p_i\in I^-(\zGamma_{\vec k,T})$, thus there
 exists a future directed causal curve $\gamma_i$ from $p_i$ to a point  $q_i\in\zG_{\vec
 k,T}$.

Consider, now, the optical function $S^+$ defined in
Appendix~\ref{sub:optical-funcS}, see \eq{Sminf}. As $\nabla S^+$
is lightlike, the function $S^+$ is strictly increasing on every
timelike curve contained within the domain of definition of $S^+$.
Suppose, first, that $\gamma_i$ is entirely contained in $\{x_n\ge
\zR\}$. Then $S^+(p_i)< S^+(q_i)$. By construction of $S^+$ we
have $S^+(q_i)=T$. On the other hand, the asymptotic behavior
\eq{Splusasym} of $S^+$ and of $\Gamma_{(1,\vec
 k'),T',\beta'}$ gives, to leading order,
$$S^+(p_i)= x^0(s)-x^n(s) +o(s)= s - k'^ns +o(s)
\to_{s\to\infty}\infty$$ except if \eq{keq} is satisfied. Thus, if
\eq{keq} does not hold, then $\gamma_i$ crosses the boundary of
the set $\{x^n\ge \zR\}$; let $r_i$ denote the first crossing
point, and let $t_i$ denote the last one. 
 Note that $\dot J^-(\zG_{\vec
 k,T};\{x^n\ge \zR\})\cap \{x^n=\zR\}$ can be represented as a graph:
 $$\dot J^-(\zGamma_{\vec
 k,T};\{x^n\ge \zR\})\cap \{x^n=\zR\}=\{x^0=s^-(x^a)\;,\ x^a\in \R^{n-1} \}\;.$$
Since $S^+\le T$ on $\dot J^-(\zGamma_{\vec
 k,T};\{x^n\ge \zR\})\cap \{x^n=\zR\}$ we have, by
 \eq{Splusasym},
\bel{slog} s^-=O(\ln (2+\rho))\;.\ee

 Somewhat similarly, $\dot J^+(p_1; \{r\ge \zR\})\cap \{x^n=\zR\}$ is a graph of a
 function $s^+$.  With our choice of $\zR$ the slopes of the light
 cones in $\Mext$ are bounded from below by one half, which
 implies
\bean s^+(x^a)&\ge&  \underbrace{x^0(s)-1+\frac {x^n(s)-\zR}2 }_I+
\underbrace{\frac 12 \sqrt{\sum_a
 (x^a-x^a(s))^2}}_{II}
 \\ &\ge & \frac 54 s + \frac 12 \sqrt{\sum_a
 (x^a-x^a(s))^2}-C\;,\eeal{rine}
 for some $s$--independent constant $C$.
 Here the contribution denoted as $I$ is obtained by following a
 coordinate line, of slope one half lying in a $x^0-x^n$ plane,
 from $p_1$ to a point $r\in \{x^n=\zR\}$, while the
 contribution denoted as $II$ is obtained from a cone of slope one
 half issued from $r$.

Now
 \beaa \rho &=&\sqrt{\sum_a
 (x^a)^2}\le \sqrt{\sum_a
 (x^a-x^a(s))^2}+ \sqrt{\sum_a
 (x^a(s))^2} \\ & \le & \sqrt{\sum_a
 (x^a-x^a(s))^2}+ \frac 12 s + C'
 \;,\eeaa
 so that
$$\rho -\frac 12s -C' \le  \sqrt{\sum_a
 (x^a-x^a(s))^2}
 \;,$$
 and \eq{rine} gives
\bel{slin} s^+(x^a)\ge  s + \frac 12 \rho-\frac 12 C'-C\;.\ee
It follows that for any future directed causal curve starting from
one of the points $p_i$ and entirely contained in $\{r\ge\zR\}$,
at each crossing point \eq{slin} holds.

Suppose that $\gamma_i$ never exits $\{r\ge \zR\}$. Then at the
last point $t_i=(x^0_i,x^a_i)$ at which $\gamma_i$ crosses
$\{x^n=\zR\}$ we must have
$$s^+(x^a_i)\le s^-(x^a_i)\;,$$
which is not possible by \eq{slog} and \eq{slin} if $s$ is chosen
sufficiently large. Thus, either \eq{keq} holds, or $\gamma_i$
enters and exits $\{r\ge \zR\}$ (perhaps more than once). Let
$\hat r_i $ denote the first
exit point. 
Let $\hat s^-$ be the graphing function of $\dot J^-(\zGamma_{\vec
k,T};\mcM)\cap \{r=\zR\}$ (exceptionally, for emphasis,  we write
$\dot J^-(\zGamma_{\vec k,T};\mcM)$ for $\dot J^-(\zGamma_{\vec
k,T})$):
$$\dot J^-(\zGamma_{\vec k,T};\mcM)\cap \{r=\zR\}=\{x^0=\hat
s^-(v)\;, v\in S^{n-1}(\zR)\subset \R^n\}\;.$$ Since $\hat r_i
=(\hat x^0_i,\hat v_i)\in J^-(\zGamma_{\vec
 k,T};\mcM)$ we have $\hat x^0_i \le \hat s^-(v_i)\le \sup \hat s^-<\infty$, the last inequality following from
  compactness and from the fact that $$ J^-(\zGamma_{\vec
 k,T};\mcM)\cap \{r=\zR\}\supset\zGamma_{\vec
 k,T}\cap \{r=\zR\}\quad \Longrightarrow \quad J^-(\zGamma_{\vec
 k,T};\mcM)\cap \{r=\zR\}\ne \emptyset\;,$$ so that $\hat s^-\not \equiv \infty$.
 However, $x^0$ is increasing along $\gamma_i$, so that $x^0_i\ge
 x^0(s)-1\to_{s\to \infty} \infty$. 
%
%
%
%
It follows that an appropriate choice of $s$ guarantees that all
the $\gamma_i$'s are entirely contained within $\{r\ge \zR\}$.
Consequently, \eq{keq} must be true.

 {}From \eq{keq} and the definition of the $p_i$'s we have $S^+(p_i)\to_{i\to\infty} T'$,  which
implies $T'\le T$. Lemma~\ref{Lmon3} shows that no point of the
form $\Gamma_{(1,\vec k),T',\beta'}(s)$, with $T'<T$ and $s$
large, can be in $\dot J^-(\zGamma_{\vec k,T})$, hence
$$ T'=T\;.$$
We have thus shown \bel{horinc}\mcN_{\vec k,T}\subset
\cup_{\beta\in \R^n}\Gamma_{(1,\vec k),T,\beta}\;.\ee Equality
ensues now from the fact that the intersections of both sets with
the region $\{x^n>\zR\}$ are smooth closed hypersurfaces.
\qed

\begin{Corollary}
\label{Clines} Each maximally extended geodesic $\Gamma$ in $\doc$
such that $r(\Gamma(s))\to_{s\to\infty}\infty$ is a line.
\end{Corollary}

\proof By Corollary~\ref{Cuniq} every such future directed
$\Gamma$ coincides with some $\Gamma_{(1,\vec k),T,\beta}$. But
Corollary~\ref{Lmon} and Proposition~\ref{Lmon2} show that each
$\Gamma_{(1,\vec k),T,\beta}$ is a line. The result for
past-directed geodesics follows by changing time orientation. \qed

Returning to the proof of Theorem~\ref{Tfoliate}, we claim that
each point $p$ in $\doc$ lies on some $\Gamma_{(1,\vec
k),T,\beta}$. Indeed, Lemma~\ref{Lcocon} with the direction
$(0,\ldots,0,1)$ rotated into
$\vec k$ provides a geodesic ray 
through $p$ with asymptotic direction $\vec k$, and the result
follows from Proposition~\ref{Puniq}.

We have thus proved that each of the distinct families of null
hypersurfaces
$$\mcN_{\vec k}:= \{\mcN_{\vec k,T}\}_{T\in\R}$$
covers $\doc$. Suppose that $\doc\ne \mcM$, then there exists a
point $$p\in I^-(\partial\hyp)\cap I^-(\Mext)\setminus \doc\subset
J^-(\Mext)$$ and, by Lemma~\ref{Lcocon}, a future directed null
half-line from $p$ extending to infinity in $\Mext$. This is,
however, impossible by Corollary~\ref{Clines}, and the proof is
complete. \qed

\noindent {\sc Proof of Theorem~\ref{Tpoormannulnew}:} By
Theorem~\ref{Tfoliate} every maximally extended geodesic is a
line. In view of the Raychaudhuri equation this is only possible
if for all null vectors $k$ we have
\bel{nullc} \mathrm{Ric}(k,k)=0\;. \ee Elementary algebra implies then that
the Ricci tensor is proportional to the metric. For future
reference we note a somewhat stronger result, which does not
assume that \eq{nullc} holds for all $k$:

\begin{Lemma}
\label{LRic}  Suppose that there exists an open set $\Omega$ of null
vectors $k$ at $p\in \mcM$ for which we have
$$R_{\mu\nu}k^\mu k^\nu =0\;.$$ Then $R_{\mu\nu}$(p) is proportional to $g_{\mu\nu}$.
\end{Lemma}

\proof Let $e^\mu$ be an ON-frame at $p$, (changing $e^0$ to
$-e^0$
 if necessary) in this frame we write $k=(|\vec k|_\delta ,\vec k)$,
 and there exists an open
cone $\mcU\subset \R^{n-1}$ of $\vec k$'s such that
\bel{eqalph} \alpha_{ij}k^ik^j + 2 \beta_i k^i |\vec k|_\delta=0\;,\ee where
$\beta_i=R_{0i}(p)$,
$\alpha_{ij}=R_{ij}(p)+R_{00}(p)\delta_{ij}$. 
Let $\vec k_0\in \mcU$, after rescaling  we can assume that $|\vec
k_0|_\delta =1$. Let $\vec m$ be any vector such that $\vec m
\cdot \vec k_0=0$, $|\vec m|_\delta =1$. Setting $\vec k = \vec
k_0 + x \vec m$\, in \eq{eqalph} one obtains an equation of the
form
$$f(x)=a+bx+cx^2+(d+ex)\sqrt{1+x^2}=0\;,$$
for all $x$ in a neighborhood of zero. As $f$ is analytic in a
neighborhood of zero, all the coefficients of its power series
there vanish, leading to
$a=b=c=d=e=0$, which is equivalent to
$$\alpha_{ij}k_0^ik_0^j =\alpha_{ij}k_0^i m^j =\alpha_{ij}m^i m^j = \beta_i
k_0^i=\beta_i m^i = 0\;.$$ Since $m$ is arbitrary we obtain
$\beta_i=0$, and $\alpha_{ij}=0$ follows by polarisation.\qed

Returning to the proof of Theorem~\ref{Tpoormannulnew},
Lemma~\ref{LRic} shows that the Ricci tensor is proportional to
the metric. By a Bianchi identity the proportionality factor must
be a constant, and asymptotic flatness shows that the constant is
zero. Therefore $(\mcM,g)$ is null geodesically complete and
vacuum, and we conclude by~\cite{Galloway:splitting}. \qed

\appendix
\section{Null geodesics extending to infinity in uniformly Schwarzschildian space-times}
\label{SA}

 In this appendix we study null geodesics in $\Mext$, as needed in the argument above.
 We start by showing existence of null geodesics $\Gamma(s)$
 with freely prescribable asymptotic direction, which can without loss of generality be assumed
 to be $e_0+e_n$, as well as freely prescribable values
 $$\beta^a:=\lim_{s\to\infty} \Gamma^a(s)\;,\quad a=1,\ldots,n-1\;.$$
 We will assume that there exist constants $C_0>0$ and $\epsilon>0$
 such that,
in the coordinates of \eq{stschw} on $\Mext$,
\beal{guS1} &|g-g_m| \le C_0r^{-1-\epsilon}\;,\quad
|\partial_\mu\left(g-g_m \right)|\le C_0r^{-2-\epsilon}\;, & \\
& |\partial_\mu\partial_\nu\left(g-g_m \right)|\le
C_0r^{-3-\epsilon}\;.\eeal{guS2} When $m=0$ and $g$ is flat in the
asymptotic region all the results established in this appendix
hold trivially, and therefore from now on we assume that $m\ne 0$.
On the other hand, for the purposes of the analysis of the
behavior of the null geodesics in this appendix, the separation of
$g$ into a Schwarzschild part and a remainder is only relevant in
dimension $n=3$, so to avoid unnecessary discussions  we assume
here that
$$0<\epsilon \le\max(1, n-3)\;.$$

 \subsection{Asymptotic behavior of null geodesics}

 Let
 \bel{alpf}\alpha=(\alpha^\mu)=(T,\beta=(\beta^a),R)\in \R\times \R^{n-1}\times \R\;,\ee $$b=|\beta|_\delta\;,$$
 we want to construct geodesics $$[0,\infty)\ni s \to x^\mu(s)\in \Mext$$
 of the form
\bel{desform}x^\mu(s)=\alpha^\mu+ sk^\mu + \underbrace{\psi^\mu_0(s)+\delta
 \psi^\mu(s)}_{\psi^\mu (s)}\;,\qquad k=(k^\mu)=(1,\vec k)\;,\ |\vec
 k|_\delta = 1\;.\ee
 The function $\psi_0^\mu$ is chosen so as
to eliminate one of the leading order terms in the geodesic
equation for $n=3$ and $\mu=0$. So we set $\psi_0^\mu\equiv 0$
unless $n=3$ and $\mu=0$, in which case we set
$$\psi_0^0(s)= 2m \ln\left
(\frac{R+s+\sqrt{(R+s)^2+b^2}}2\right)\;,$$ so that
$$\dot \psi_0^0(s) = \frac {2m} {\sqrt{(R+s)^2+b^2}}\;.$$

 \begin{Proposition}
\label{Pexg2} Let $n\ge 3$ and let $g$ satisfy \eq{guS1}-\eq{guS2}
on $\Mext$ with some $0< \epsilon \le \max(1, n-3)$. There exist
constants $\zR$ and $\zC$ such that for every $\vec
k\in S^{n-1}\subset \R^{n}$, 
$(T,\beta) \in \R\times \R^{n-1}$ and $R\ge \zR$ there exists an
affinely parameterized null geodesic of the form
\eq{alpf}-\eq{desform} satisfying
 \bel{decest2}|\delta \psi|+
\sqrt{(R+s)^2+|\beta|_\delta^2}\; |\delta \dot \psi| \le \frac
{\zC} {({(R+s)^2+|\beta|_\delta^2} )^{\epsilon/2}}\;.\ee
\end{Proposition}

\begin{Remark}
 \label{Rexg2}If the metric has a smooth, or polyhomogeneous, conformal
completion, one immediately obtains a polyhomogeneous expansion
for $\delta \psi$ in terms of $s$.
\end{Remark}

 \proof
$x^\mu(s)$ will be an affinely parameterized geodesic if and only
if $$\frac {d^2 \delta \psi^\mu}{ds^2} = F^\mu ( \delta \psi,
\delta \dot \psi ,s)\;,$$ where a dot denotes an $s$--derivative,
and
$$ F^\mu ( \chi, \lambda,s)= -\left(\Gamma^\mu_{\nu\rho}\circ(\alpha+ sk
+ \psi_0+\chi)\right)(k^\nu + \dot \psi_0^\nu+\lambda^\nu) (k^\rho
+ \dot \psi_0^\rho+\lambda^\rho) - \frac {d^2
\psi^\mu_0}{ds^2}\;.$$ {}From \eq{stschw} and \eq{guS1}-\eq{guS2}
we have $\Gamma^\mu_{\nu\rho}=O(r^{-2-\epsilon})$ except for
\beal{Gam1} & \Gamma^k_{00}=\Gamma^0_{k0}=\Gamma^0_{0k} = \frac
{(n-2)m}{r^{n-1}} \frac {x^k}r +O(r^{-2-\epsilon})\;, & \\
& \Gamma^k_{ij}=\frac m {r^{n-1}}\left(\delta_{ij} \frac {x^k}r
-\delta_{jk} \frac {x^i}r - \delta_{ik} \frac {x^j}r \right)+
O(r^{-2-\epsilon})\;.& \eeal{Gam2} We will write $\delta
\Gamma^\mu_{\nu\rho}$ for all the remainder terms not explicitly
listed above.

 In dimension three one  has
\bean F^0 & = & -\frac {2 m}{r^{2}}\left(\Big(\sum_i \frac
{x^i}{r}(1+\dot\psi_0^0+ \lambda^0)(k^i+
\lambda^i)\Big)-\frac{r^{2}(R+s)}{((R+s)^2+b^2)^{3/2}}\right)
\\ && \label{F0}-\delta\Gamma^0_{\mu\nu} (k^\mu+\dot \psi_0^\mu+\lambda^\mu) (k^\nu+\dot \psi_0^\nu+\lambda^\nu)\;, \\
F^i & = & -\frac m {r^{2}}\Bigg(\frac {x^i}r \left(|\vec k +
\lambda|_\delta^2  + (1+\dot \psi_0^0+\lambda^0)^2 \right)
\nonumber   - 2(k^i  +\lambda^i)\Big(\sum_j \frac {x^j}r(k^j +
\lambda^j) \Big)\Bigg) \nonumber
\\ && -\delta\Gamma^i_{\mu\nu}
(k^\mu+\dot \psi_0^\mu+\lambda^\mu) (k^\nu+\dot
\psi_0^\nu+\lambda^\nu)\;.\eeal{Fi} Here, and in all the
calculations that follow, $x^\mu$ has to be replaced by
$\alpha^\mu+ sk^\mu + \psi^\mu_0 + \chi^\mu$.

 On the other hand,
in higher dimension it will suffice to estimate directly
from
$$F^\alpha =-\Gamma^\alpha_{\mu\nu} (k^\mu+\lambda^\mu) (k^\nu+\lambda^\nu)\;.$$
We choose $$0<\sigma<\epsilon\;,$$ and we let $X\subset
C([0,\infty))\times C([0,\infty))$ denote the set of pairs of maps
into $\R^{n+1}$ such that the following norm is finite:
\bel{secnorm}\|(\chi,\lambda)\|_{R,b}= \|(R+s+b)^{\sigma
}\chi\|_{L^\infty([0,\infty))} +\frac{2 \|(R+s+b)^{1+\sigma
}\lambda\|_{L^\infty([0,\infty))}}{1+\sigma  } \;.\ee
%
Let $B(1)\subset X $ be the unit closed ball in $X $ and define
the map \bel{mcTmap} B(1)\ni (\chi,\lambda) \to
\mcT(\chi,\lambda)=\left (-\int_s^\infty \lambda(u)du\;,\
-\int_s^\infty F^\mu(\chi,\lambda,u)du\right)\;.\ee Clearly a
fixed point $(\chi_0,\lambda_0)$ of $\mcT$ provides an affinely
parameterized geodesic of the form \eq{desform}, with $|\delta
\psi(s)|=|\chi_0|\le  (R+s+b)^{-\sigma }$, and with
\bel{asympt} \dot x^\mu \to_{s\to\infty} k^\mu\;. \ee In order to
check that $\mcT$ does indeed have a fixed point, note first the
elementary inequality: for $R+s\ge 0$, $b\ge 0$,
\bel{elineq}\sqrt{(R+s)^2+b^2}\le R+s+b\le
\sqrt{2}\sqrt{(R+s)^2+b^2}\;.\ee It follows that for
$(\chi,\lambda)\in B(1)$ and for $R$ sufficiently large we have
\bel{labelproblem2} \frac 12  (R+s+b) \le r \le 2 (R+s+b)\;, \ee
\beal{rhocont3}|\rho - b| &\le& \frac C
{(R+s+b)^{\sigma }}\;,\\
\label{rhocont4}|x^n-R-s|=|\chi^n(s)| & \le &  \frac 1
{(R+s+b)^{\sigma }}\;.\eea This implies
\bean |r - \sqrt{(R+s)^2+b^2}|  & = & \left|\frac {r^2 -  (R+s)^2-b^2}{r
+ \sqrt{(R+s)^2+b^2}}\right| \\\nonumber & = & \left|\frac {(x^n -
(R+s))(x^n +  (R+s))+ (\rho-b)(\rho+b)}{r +
\sqrt{(R+s)^2+b^2}}\right|
\\& \le &  \frac {1+C} {(R+s+b)^{\sigma}}\;.\eeal{rcont2}
\Eq{labelproblem2} immediately gives an estimate
$$|\delta\Gamma^\sigma_{\mu\nu}
(k^\mu+\dot \psi_0^\mu+\lambda^\mu) (k^\nu+\dot
\psi_0^\nu+\lambda^\nu)|\le \frac {C_2} {(R+s+b)^{2+\epsilon}}
\;.$$  Writing \beaa F^\mu(\chi,\lambda)& = & \chi \int _0^1
\frac{\partial F^\mu}{\partial \chi} (s\chi,\lambda)ds +\lambda
\int_0^1 \frac{\partial F^\mu}{\partial \lambda}(0,s\lambda)ds +
F^\mu(0,0)\;,\eeaa with a little more work a similar estimate is
obtained for the whole of $F^\mu$ using \eq{rhocont3}-\eq{rcont2}.
This leads to
$$\left|\int_s^\infty
F^\mu(\chi,\lambda,u)du\right| \le C_3 (R+s+b)^{-1-\epsilon}\le
C_3 R^{\sigma-\epsilon   } (R+s+b)^{-(1+\sigma) }\;,$$ for some
$C_3=C_3(C_1,|m|,\epsilon)$, with $C_3$ also depending on the
constant dominating the error terms in \eq{guS1}. Increasing $R_1$
if necessary one finds that $\mcT$ maps $B(1)$ into $B(1)$ for all
$R\ge R_1$.

 One similarly checks that the map $\mcT$ defined in
\eq{mcTmap}
 is a contraction for $R$ sufficiently large, uniformly in
 $b$, and the existence of a geodesic of the form \eq{desform}, with
$\delta\psi(s)\to_{s\to\infty}0$, ensues. 
 Once we know that the solution just described exists, it is
straightforward to show, using the geodesic equation and the
estimates already available, that $\delta \psi$ satisfies
\eq{decest2}. \qed

We will need to know that all geodesics extending to infinity are
as above:

 \begin{Proposition}
\label{Puniq} Let $g$ satisfy \eq{guS1}-\eq{guS2} on $\Mext$. Then
every null geodesic $\Gamma(s)$ such that
\bel{extend} r(\Gamma(s))\to_{s\to\infty}\infty\ee satisfies \eq{alpf}-\eq{decest2}.
\end{Proposition}

\proof The existence of $k^\mu$ is provided by Proposition~B.1 in
\cite{ChmassCMP}. {}From that last proposition one also has a
uniform bound on $dx^\mu/ds$, as well as the estimate
\eq{labelproblem2}, and the remaining claims easily follow by
inspection of the geodesic equation. \qed

\begin{Corollary}
\label{Cuniq} The map which to every maximally extended, null
geodesic satisfying \eq{extend} assigns $k$, $\beta$ and $T$ is
bijective.
\end{Corollary}

\proof It remains to prove injectivity. But
Proposition~\ref{Puniq} implies that geodesics satisfying
\eq{extend} are solutions of the fixed point problem considered in
the proof of Proposition~\ref{Pexg2}, and are therefore unique.
\qed

Let $\Gamma_{k,T,\beta}(s)$ denote the unique affinely
parameterized maximally extended geodesic provided by
Corollary~\ref{Cuniq}. The differentiability properties of
$\Gamma_{k,T,\beta}$ with respect to $k$, $T$ and $\beta$ are best
studied by considering Jacobi fields along $\Gamma$. The method of
proof of Proposition~\ref{Pexg2} applies to give:

\begin{Proposition}
\label{PJacobi2} Let $\Gamma:[0,\infty)$ be a null affinely
parameterized geodesic satisfying \eq{extend}. Then for every $A,
B\in \R^{n+1}$ there exists a solution $Z$ of the Jacobi equation,
\bel{tenc6C} \mbox{$\displaystyle\frac {D^2Z}{ds^2}(s) = \mathrm{Riem}(\dot \gamma, Z)\dot \gamma$}\;,\ee
where $\mathrm{Riem}$ denotes the Riemann tensor, such that
\bel{Jacf} Z^\mu = \left\{%
\begin{array}{ll}
    (R+s) A^\mu + O(\ln (R+s+\sqrt{(R+s)^2+b^2})), & \hbox{$A^\mu \ne 0$, $n=3$;} \\
    (R+s) A^\mu + O((R+s+b)^{-\epsilon}), & \hbox{$A^\mu \ne 0$, $n>3$;} \\
    B^\mu + O((R+s+b)^{-(1+\epsilon)}), & \hbox{$A^\mu=0$,} \\
\end{array}%
\right.\ee with the error terms being uniform in
$b$.\end{Proposition}

Similarly to Proposition~\ref{Puniq},  every Jacobi field along
$\Gamma$ as in Proposition~\ref{PJacobi2} is a linear combination
of solutions of the form \eq{Jacf}; this is proved by usual
techniques.

It is standard to show now
\begin{Proposition}
\label{Pcont} The map
$$S^{n-1}\times\R\times \R^{n-1}\times \R \supset \mcU \ni (\vec k,
T,\beta,s)\to \Gamma_{(1,\vec k),T,\beta}(s)$$ is differentiable
on the open set $\mcU$ on which it is defined.
\end{Proposition}

 Fix $\beta, R$, with $R\ge R_1$, and let \bel{Phimap} \Phi:\R\to\R\ee be the
map which to $T\in \R$ assigns $x^0(0)$, the coordinate time at
which the null geodesic, as constructed in the proof of
Proposition~\ref{Pexg2}, intersects $\{x^n=R\}$.
Proposition~\ref{PJacobi2} (with $A=0$) and standard ODE
considerations show that $\Phi$ is differentiable,  with strictly
positive derivative (perhaps increasing $R_1$ if necessary) and
hence strictly increasing, in particular $\tau$ is continuous. As
the constant $\zC$ in \eq{decest2} is uniform in $T$, one clearly
has
$$\tau(T)\to_{T\to \pm \infty}\pm \infty\;.$$ It follows that
$\Phi$ is a smooth bijection from $\R$ to itself.

\subsection{The optical functions $S^\pm$}
\label{sub:optical-funcS} There exists $\zR_1$ such that, using
the implicit function theorem, one can define a function
\bel{Sminf} S^+: \{x^n \ge \zR_1\}\to \R\ee by assigning
$T$ to each point lying on $\Gamma_{k,T,\beta}$. Then $S^+$ is
differentiable, with null gradient, and satisfies
\bel{Splusasym} S^+= x^0-x^n+\zR- \left\{%
\begin{array}{ll}
    2m \ln\left
(\frac{x^n+\sqrt{(x^n)^2+\rho^2}}2\right)+o(1), & \hbox{$n=3$;} \\
    o(1), & \hbox{$n\ge 4$,} \\
\end{array}%
\right.    \ee with the error term tending to zero as $r(x)$ goes
to infinity, uniformly with respect to $(x^1,\ldots,x^{n-1})\in
\R^{n-1}$.  We give some details of the construction, because the
function $S^+$ plays an important role in Section~\ref{Sproof};
furthermore, the proof of existence of $S^+$ (and its time-reverse
counterpart $S^-$), with the required properties, is the missing
step in the argument in~\cite{PenroseSorkinWoolgar}. Let
$$(y^\mu)=(T,\beta,s)\;,$$ then \eq{desform} with
$k^\mu=(1,0,\ldots,0,1)$ and $R=\zR$ (as given by
Proposition~\ref{Pexg2}) defines a map $x^\mu(y^\alpha)$. We have
\begin{enumerate}
\item $\partial x^\mu/\partial y^0$ at the point $x^\mu(y^\alpha)$ is given
by the value of the Jacobi field $Z^\mu$ which asymptotes to
$\delta^\mu_0$ as $s=y^n$ tends to infinity; from the proof of
Proposition~\ref{PJacobi2} we thus have
$$\frac{\partial x^\mu}{\partial y^0} = \delta^\mu_0+
O\Big(\Big(\zR+y^n+\sqrt{\sum_a (y^a)^2}\Big)^{-1}\Big)\;.$$
\item Similarly $\partial x^\mu/\partial y^a$ at $x^\mu(y^\alpha)$ is given
by the value of the Jacobi field $Z^\mu$ which asymptotes to
$\delta^\mu_a$ as $s=y^n$ tends to infinity; from the proof of
Proposition~\ref{PJacobi2} one finds
$$\frac{\partial x^\mu}{\partial y^a} = \delta^\mu_
a+ O\Big(\Big(\zR+y^n+\sqrt{\sum_a (y^a)^2}\Big)^{-1}\Big)\;.$$
\item Directly from its definition, $\partial x^\mu/\partial y^n= \partial x^\mu /\partial s$
equals $k^\mu + d\psi^\mu/ds$, so that
$$\frac{ \partial x^\mu }{\partial y^n} = (1,0,\ldots,0,1) + O\Big(\Big(\zR+y^n+\sqrt{\sum_a
(y^a)^2}\Big)^{-1}\Big)\;.$$
\end{enumerate}
Increasing $\zR$ if necessary, it follows that $y^\mu\to x^\mu$ is
a smooth local diffeomorphism for $r(y)\ge \zR$.

To prove bijectivity with the image we need the following,
certainly well known, result:

\begin{Proposition}
\label{Pwk} Let $\mcO\subset \R^n$ be an open convex set and let
$\Phi:\mcO\to\R^n$ be continuously differentiable. Then there
exists $\epsilon=\epsilon(n)>0$ such that if
\bel{epsicon}|\frac{\partial \Phi^\mu}{\partial
y^\nu}-\delta^\mu_\nu|<\epsilon\;,\ee then $\Phi$ is injective. If
$\mcO=\R^n$ then $\Phi(\mcO)=\R^n$.
\end{Proposition}

\proof The calculation follows those
in~\cite[Theorem~3.1]{Howard:ift}. We have \beaa
|\Phi(x)-\Phi(y)-(x-y)|& = &
\Big|\int_0^1\Big(\Phi'(tx+(1-t)y)-\id\Big)(x-y)dt \Big|
\\ &\le& C(n)\epsilon |x-y|\;.\eeaa
This implies
\beaa |x-y| &\le&|\Phi(x)-\Phi(y)| +|\Phi(x)-\Phi(y)-(x-y)|
\\ &\le & |\Phi(x)-\Phi(y)|+ C(n)\epsilon
|x-y| \;.\eeaa If $C(n)\epsilon<1$ we obtain \bel{goeq}|x-y| \le
\frac 1 {1-C(n)\epsilon}|\Phi(x)-\Phi(y)|\;,\ee and injectivity
follows. By the implicit function theorem $\Phi$ is open. It
further follows from \eq{goeq} that if the sequence
$(\Phi(x_i))_{i\in\N}$ is Cauchy, then so is the sequence
$(x_i)_{i\in\N}$. This shows that if $\mcO=\R^n$ then $\Phi(\mcO)$
is closed, proving surjectivity. \qed

For $R$ large enough consider the map $\Phi_R$ which to
$(T,\beta)\in \R^{n-1}$ assigns $x^\mu(s)$, where $s=s(T,\beta)$
is chosen so that $x^n(s)=R$. By arguments similar to the ones
already given one finds that $\Phi_R$ satisfies the hypotheses of
Proposition~\ref{Pwk} for $R\ge \zR_1$, increasing $\zR_1$ if
necessary. Hence the $\Phi_R$'s are bijective for $R\ge \zR_1$,
which implies that every point in the region $\{x^n\ge \zR_1\}$
lies on \emph{precisely one} null geodesic with asymptotic
velocity $(1,0,\ldots,0,1)$. It follows that the map $y^\mu \to
x^\mu(y^\alpha)$ is injective onto $\{x^n\ge \zR_1\}$, and
therefore is a global diffeomorphism from some set  $\mcU\subset
[0,\infty)\times \R^{n}$ to $\{x^n\ge \zR_1\}$. Inverting this map
one obtains smooth functions $y^\mu(x^\alpha)$ on $\{x^n\ge
\zR_1\}$. The function $S^+$ is then set to be equal to
$y^0(x^\mu)$. \Eq{Splusasym} is obtained by eliminating $s$ and
$\beta$ from the definition
$$S^+(x^\mu)=T(x^\mu)=x^0(s)-s-\psi^0(s)$$ using equations
\eq{desform}.

As already pointed out, the function $S^-$ is obtained by a
time-reverse of the construction just carried out.

Because the estimates above are uniform in $r(y)$ for $r(y)$
large, it should be clear that the domain of definition of $S^+$
can be extended to a region of the form $\{r\ge \zR_1\;, x^n\ge
0\}$, increasing $\zR_1$ if necessary. Similarly $S^-$ can be
defined on a region of the form $\{r(x)\ge \zR_1\;, x^n\le 0\}$.
Those extensions are relevant to the original
Penrose-Sorkin-Woolgar argument.

\bigskip

\noindent{\sc Acknowledgements:} I am grateful to Gregory Galloway
and Eric Woolgar for useful discussions, and to Malcolm~MacCallum
for bibliographical advice.

\bibliographystyle{amsplain}
\bibliography{../references/bartnik,%
../references/myGR,%
../references/newbiblio,%
../references/newbib,%
../references/reffile,%
../references/bibl,%
../references/Energy,%
../references/hip_bib,%
../references/netbiblio}
\end{document}